\documentclass{ws-procs975x65}

\begin{document}

\title{The ``Fireshell'' model in the Swift era}

\author{C.L. Bianco,$^{1,2}$ R. Ruffini,$^{1,2,3}$}

\address{
$^1$ ICRANet and ICRA, Piazzale della Repubblica 10, I-65122 Pescara, Italy.\\
$^2$ Dip. di Fisica, Universit\`a di Roma ``La Sapienza'', Piazzale Aldo Moro 5, I-00185 Roma, Italy.\\
$^3$ ICRANet, Universit\'e de Nice Sophia Antipolis, Grand Ch\^ateau, BP 2135, 28, avenue de Valrose, 06103 NICE CEDEX 2, France.\\
E-mails: bianco@icra.it, ruffini@icra.it.
}

\begin{abstract}
We here re-examine the validity of the constant-index power-law relation between the fireshell Lorentz gamma factor and its radial coordinate, usually adopted in the current Gamma-Ray Burst (GRB) literature on the grounds of an ``ultrarelativistic'' approximation. Such expressions are found to be mathematically correct but only approximately valid in a very limited range of the physical and astrophysical parameters and in an asymptotic regime which is reached only for a very short time, if any.
\end{abstract}

\bodymatter

\section{Introduction}

The consensus has been reached that the afterglow emission originates from a relativistic thin shell of baryonic matter propagating in the CircumBurst Medium (CBM) and that its description can be obtained from the relativistic conservation laws of energy and momentum. In both our approach and in the other ones in the current literature (see e.g. Refs.~\refcite{1999PhR...314..575P,1999ApJ...512..699C,2005ApJ...620L..23B,2005ApJ...633L..13B,2007AIPC..910...55R}) such conservations laws are used. The main difference is that in the current literature an ultra-relativistic approximation, following the Blandford \& McKee self-similar solution,\cite{1976PhFl...19.1130B} is widely adopted, leading to a simple constant-index power-law relations between the Lorentz $\gamma$ factor of the optically thin ``fireshell'' and its radius:
\begin{equation}
\gamma\propto r^{-a}\, ,
\label{gr0}
\end{equation}
with $a=3$ in the fully radiative case and $a=3/2$ in the adiabatic case.\cite{1999PhR...314..575P,2005ApJ...633L..13B} On the contrary, we use the exact solutions of the equations of motion of the fireshell.\cite{2004ApJ...605L...1B,2005ApJ...620L..23B,2005ApJ...633L..13B,2006ApJ...644L.105B,2007AIPC..910...55R}

\section{Exact vs. approximate solutions in the Swift era}

A detailed comparison between the equations used in the two approaches has been presented in Refs.~\refcite{2004ApJ...605L...1B,2005ApJ...620L..23B,2005ApJ...633L..13B,2006ApJ...644L.105B}. In particular, in Ref.~\refcite{2005ApJ...633L..13B} it is shown that the regime represented in Eq.(\ref{gr0}) is reached only asymptotically when
\begin{equation}
\gamma_\circ \gg \gamma \gg 1
\label{cond1}
\end{equation}
in the fully radiative regime and
\begin{equation}
\gamma_\circ^2 \gg \gamma^2 \gg 1
\label{cond2}
\end{equation}
in the adiabatic regime, where $\gamma_\circ$ the initial Lorentz gamma factor of the optically thin fireshell.

\begin{figure}
\centering
\includegraphics[width=0.75\hsize,clip]{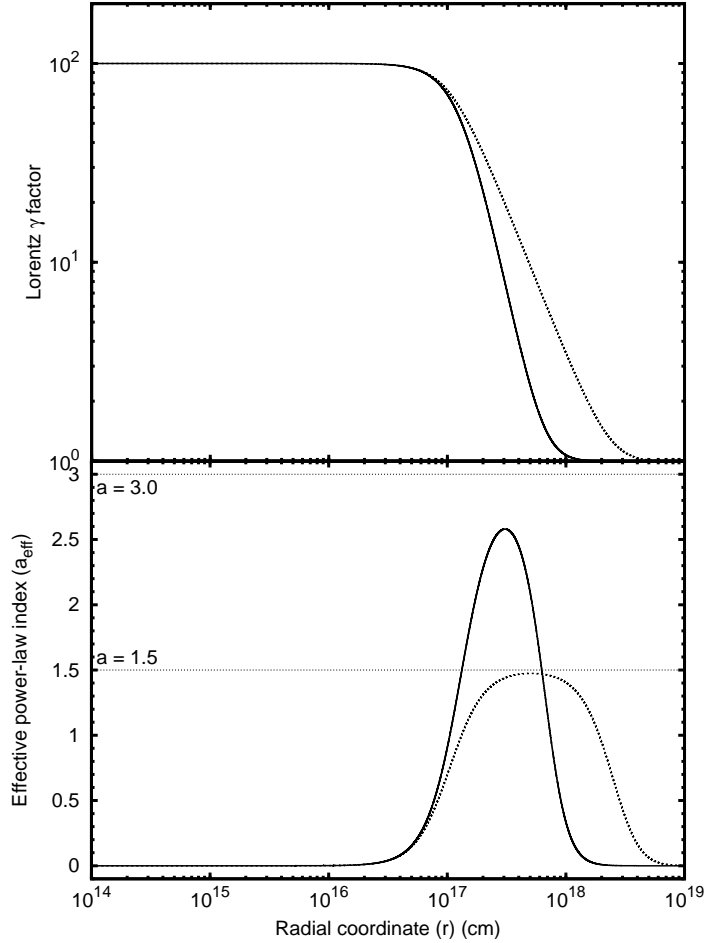}
\caption{In the upper panel, the analytic behavior of the Lorentz $\gamma$ factor during the afterglow era is plotted versus the radial coordinate of the expanding optically thin fireshell in the fully radiative case (solid line) and in the adiabatic case (dotted line) starting from $\gamma_\circ = 10^2$ and the same initial conditions as GRB 991216\cite{2005ApJ...633L..13B}. In the lower panel are plotted the corresponding values of the ``effective'' power-law index $a_{eff}$ (see Eq.(\ref{eff_a})), which is clearly not constant, is highly varying and systematically lower than the constant values $3$ and $3/2$ purported in the current literature (horizontal thin dotted lines).}
\label{gdir_a_comp_rad-ad_mg11}
\end{figure}

In Fig.~\ref{gdir_a_comp_rad-ad_mg11} we show the differences between the two approaches. In the upper panel there are plotted the exact solutions for the fireshell dynamics in the fully radiative and adiabatic cases. In the lower panel we plot the corresponding ``effective'' power-law index $a_{eff}$, defined as the index of the power-law tangent to the exact solution:\cite{2005ApJ...633L..13B}
\begin{equation}
a_{eff} = - \frac{d\ln\gamma}{d\ln r}\, .
\label{eff_a}
\end{equation}
Such an ``effective'' power-law index of the exact solution smoothly varies from $0$ to a maximum value which is always smaller than $3$ or $3/2$, in the fully radiative and adiabatic cases respectively, and finally decreases back to $0$ (see Fig.~\ref{gdir_a_comp_rad-ad_mg11}). 

Thanks to the \emph{Swift} satellite\cite{2004ApJ...611.1005G}, we have now for many GRBs almost gapless multi-wavelength light curves from the beginning of the prompt emission (which in our model coincides with the peak of the afterglow, see Refs.~\refcite{2001ApJ...555L.113R,2006ApJ...645L.109R,2007A&A...471L..29D,2007A&A...474L..13B,bianco_ita-sino}) all the way to the latest afterglow phases. In the interpretation of such gapless data it is therefore crucial to use the exact solution for the fireshell dynamics.

\end{document}